\documentclass[english,12pt]{article}
\usepackage{array}
\usepackage{graphicx}
\usepackage{amssymb}
\usepackage{amsmath}
\usepackage{multirow}
\usepackage{prettyref}
\usepackage{babel}
\usepackage{units}
\usepackage[latin1]{inputenc}
\usepackage{amsfonts}
\usepackage{amssymb}
\usepackage{babel}
\usepackage{subcaption}
\usepackage{cite}
\def\@fmsl@sh#1#2#3{\m@th\ooalign{$\hfil#1\mkern#2/\hfil$\crcr$#1#3$}}
 \def\eq#1\en{\begin{equation}#1\end{equation}}
\def\s[#1,#2]{[#1\stackrel{\star}{,}#2]}
\def\sx[#1,#2]{[#1\stackrel{\star_{x}}{,}#2]}

\newcommand{\nc}{\newcommand}
\nc{\beq}{\begin{equation}}
\nc{\eeq}{\end{equation}}
\nc{\beqa}{\begin{eqnarray}}
\nc{\eeqa}{\end{eqnarray}}

\def\bc{\begin{center}}
\def\ec{\end{center}}

\def\to{\rightarrow}

\def\gsim{\mathrel{\mathpalette\atversim>}}

\def\bc{\begin{center}}
\def\ec{\end{center}}

\def\gsim{\mathrel{\rlap{\lower4pt\hbox{\hskip1pt$\sim$}}

    \raise1pt\hbox{$>$}}}       

\def\gsim{\mathrel{\rlap{\lower4pt\hbox{\hskip1pt$\sim$}}
    \raise1pt\hbox{$>$}}}       

\begin{document}
\makeatletter
\def\fmslash{\@ifnextchar[{\fmsl@sh}{\fmsl@sh[0mu]}}
\def\fmsl@sh[#1]#2{%
  \mathchoice
    {\@fmsl@sh\displaystyle{#1}{#2}}%
    {\@fmsl@sh\textstyle{#1}{#2}}%
    {\@fmsl@sh\scriptstyle{#1}{#2}}%
    {\@fmsl@sh\scriptscriptstyle{#1}{#2}}}
\def\@fmsl@sh#1#2#3{\m@th\ooalign{$\hfil#1\mkern#2/\hfil$\crcr$#1#3$}}
\makeatother

\thispagestyle{empty}
\begin{titlepage}
\boldmath
\begin{center}
  \Large {\bf Spacetime instability due to quantum gravity}
    \end{center}
\unboldmath
\vspace{0.2cm}
\begin{center}
{{\large Iber\^e Kuntz\footnote{ibere.kuntz@ufabc.edu.br}, Rold\~ao da Rocha\footnote{roldao.rocha@ufabc.edu.br}}}
 \end{center}
\begin{center}
{\sl Federal University of ABC, Center of Mathematics,  Santo Andr\'e, Brazil
}
\end{center}
\vspace{5cm}
\begin{abstract}
\noindent We show that quantum gravity yields exponentially growing gravitational waves. Without a mechanism to stop these modes from growing, the universe would go through a gravitational collapse. For Minkowski background, we propose a solution by choosing an integration contour in Fourier space that does not enclose the problematic modes, thus preventing them from showing up in the effective theory. It turns out that this is only possible when the modes are removed altogether. For an expanding universe, we argue that the runaway modes can be managed accordingly to the dynamics of the Hubble constant, leading to important implications for astrophysics.
\end{abstract}  
\end{titlepage}



\newpage

\section{Introduction}
\quad\, Of all fundamental interactions, gravity is probably the most challenging one. Even though it is the oldest of the forces and the only one that is part of everyone's daily lives, it is still the one lacking a full quantum treatment. Attempts of quantizing gravity has led to inumerous difficulties over the years, with partial success obtained only in the low-energy regime.

One of the issues that remains unsolved concerns the vacuum stability. In the semi-classical limit, where the graviton is not quantized, instabilities are plagued in the form of exponentially growing fields. It has been shown that massive and massless scalar fields coupled to classical gravity renders instability in the Minkowski spacetime \cite{Horowitz:1980fj,RandjbarDaemi:1981wd}. Quantum general relativity at finite temperatures was also shown to produce instabilities in the flat spacetime \cite{Gross:1982cv}. Further studies were performed in \cite{Biran:1983ip,Anderson:2002fk,Hu:2004wq}. Similar studies were also made for the de Sitter spacetime \cite{Polyakov:2007mm,Polyakov:2009nq,Polyakov:2012uc,Anderson:2013zia,Anderson:2013ila,PerezNadal:2007ez,Ho:2015bua}.

In this brief paper, we will investigate one-loop corrections to general relativity using effective field theory techniques and we will show that linear perturbations around Minkowski and de Sitter spacetimes render exponentially growing solutions of the wave equation, indicating that these backgrounds suffer from instabilities due to quantum gravity. Our approach builds up on the literature in two different ways as: i) we take into account the quantization of the graviton as well as of massless matter fields and ii) we show that instabilities arise even for self-interacting gravitons, with no matter present. We also propose a solution whereby we pick up a suitable contour of integration without including the runaway modes to the theory.

After integrating out the graviton fluctuations with the background field method, one obtains both local and non-local contributions to the quantum gravitational effective action \cite{Barvinsky:1987uw,Donoghue:2014yha}. However, to make the presentation of this article simpler and without loss of generality, we will focus only on the non-local contribution to the Einstein-Hilbert action. The only effect of the inclusion of the local part is to change the position of the roots \eqref{eq:q1} and \eqref{eq:q2} below, without changing their complex nature \cite{Calmet:2017rxl}. The quantum effective action then reads
\begin{equation}
\Gamma = \int\mathrm{d}^4x\sqrt{-g}\left(\frac{M_p^2}{2}R - c_1 R\log\frac{\Box}{\mu^2}R - c_2 R_{\mu\nu}\log\frac{\Box}{\mu^2}R^{\mu\nu} - c_3 R_{\mu\nu\rho\sigma}\log\frac{\Box}{\mu^2}R^{\mu\nu\rho\sigma}\right),
\label{eq:nl}
\end{equation}
where $M_p = (8\pi G)^{-1/2}$ is the reduced Planck mass, $G$ is the Newton's constant, $\mu$ is the renormalization scale and the kernel $R$ denotes the Riemann tensor and its contractions (Ricci tensor and Ricci scalar) depending on the number of indices it carries. The signature $(+---)$ will be adopted. The coefficients $c_i$ are predictions of the infra-red theory and depend only on the field content under consideration (see Table 1 in \cite{Donoghue:2014yha} for their precise values).

To study the stability of a background, we will linearize \eqref{eq:nl} using $g_{\mu\nu}=\bar g_{\mu\nu}+h_{\mu\nu}$, where $h_{\mu\nu}$ stands for perturbations around the background metric $\bar g_{\mu\nu}$. It is the behavior of $h_{\mu\nu}$ that will tell us about the spacetime stability. After linearization, the equation of motion obtained from \eqref{eq:nl} is \cite{Kuntz:2017pjd}
\begin{equation}
F(\Box_{\bar g})h_{\mu\nu} = 0,
\label{eq:eom}
\end{equation}
where
\begin{equation}
F(\Box_{\bar g})\equiv\Box_{\bar g} + \frac{N G}{120\pi}\log\left(\frac{\Box_{\bar g}}{\mu^2}\right)\Box^2_{\bar g}
\label{eq:op}
\end{equation}
and $\Box_{\bar g} = \bar g^{\mu\nu}\bar\nabla_\mu\bar\nabla_\nu$. Here $N = N_s+6N_f+12N_V+42$ and $N_s$, $N_f$ and $N_V$ denote the number of scalar, spinor and vector fields in the theory, respectively. The numerical value $42$ represents the graviton contribution \cite{Donoghue:2014yha}. Symbols with bar indicate that they are defined using the background metric $\bar g_{\mu\nu}$.

This paper is organized as follows. In Section \ref{sec:mink}, we solve the equation of motion \eqref{eq:eom} for $h_{\mu\nu}$ around Minkowski background $\eta_{\mu\nu}$ and we show that the most general solution in empty space contains both damped and exponentially growing solutions. The latter signs an instability of Minkowski spacetime. We argue that this issue can only be solved by choosing a contour of the fourier transform that does not enclose any mode. This implies that the only possible solution is the trivial one $h_{\mu\nu}=0$. In Section \ref{sec:sitter}, we solve the equation of motion for a de Sitter background and we also find exponentially growing modes. In this case, however, an alternative solution exists as the growing solutions are negligible for a large Hubble constant $H\gg \frac{M_p}{\sqrt{N}}$. We draw the conclusions in Section \ref{sec:conc}.

\section{Stability of Minkowski spacetime}
\label{sec:mink}
In this section we take $\bar g_{\mu\nu}=\eta_{\mu\nu}$. Let us write $h_{\mu\nu}$ in terms of its Fourier modes
\begin{equation}
h_{\mu\nu}(x) = \oint_\mathcal{C}\mathrm{d}^4q\, e^{-iqx}\tilde h_{\mu\nu}(q),
\label{eq:fourier}
\end{equation}
where $\mathcal{C}$ stands for a contour of integration in the complex plane to be chosen. As we are working within the realm of effective field theory, we are allowed to choose the contour $\mathcal C$ as we wish. It is precisely this freedom of picking up the integration contour that will permit us to remove the instabilities from the theory. The pseudo-differential operator \eqref{eq:op} acts on \eqref{eq:fourier} as \cite{Barnaby:2007ve}
\begin{equation}
F(\Box_\eta)h_{\mu\nu} = \oint_\mathcal{C}\mathrm{d}^4q\, e^{-iqx} F(-q^2) \tilde h_{\mu\nu}(q),
\end{equation}
where
\begin{equation}
F(-q^2) = -q^2\left[1-\frac{N G}{120\pi} q^2\log\left(-\frac{q^2}{\mu^2}\right)\right].
\end{equation}
Thus the equation of motion \eqref{eq:eom} becomes
\begin{equation}
\oint_\mathcal{C}\mathrm{d}^4q\, e^{-iqx}q^2\left[1-\frac{N G}{120\pi} q^2\log\left(-\frac{q^2}{\mu^2}\right)\right]\tilde h_{\mu\nu}(q) = 0.
\label{eq:eomq}
\end{equation}
To solve this equation, we make use of Cauchy's integral theorem which states that a contour integral vanishes if its integrand is analytic everywhere inside $\mathcal{C}$. Thus the solution $\tilde h_{\mu\nu}(q)$ must only develop poles that are canceled by the zeros of $F(-q^2)$, which are given by \cite{Calmet:2014gya}
\begin{align}
\label{eq:q0}
q_0 &= 0,\\
\label{eq:q1}
q_1^\pm &=\pm \sqrt{\frac{120\pi}{NG}\frac{1}{W\left(\frac{-120\pi}{\mu^2 N G}\right)}},\\
\label{eq:q2}
q_2^\pm &=\pm \sqrt{(q_1^2)^*},
\end{align}
where $W(x)$ denotes the Lambert W-function. The exact position of these zeros depends on the renormalization scale $\mu$. For practical purposes, we pick up $\mu$ so that the argument of the Lambert W-function is $-1$:
\begin{equation}
\mu = \sqrt{\frac{120\pi}{NG}}.
\label{eq:scale}
\end{equation}
The important point is that there will always be zeros with positive and negative imaginary parts.

Naively, one would think that all these zeros of $F(-q^2)$ lead to wave solutions in the position space. However, there is a subtlety regarding the contour of integration $\mathcal{C}$ of the Fourier transform that we shall now explain. The most natural contour, from a mathematical viewpoint, would be one that encloses all of the zeros of $F(-q^2)$. However, the pole at the origin coincides with the branch point of the logarithm. To make the logarithm a single-valued function, we must take a branch cut connecting the origin to a point at infinity. Consequently, it is impossible to find a contour that encloses the pole $q_0 = 0$ without crossing the branch cut. In fact, while the branch cut is artificial and can be chosen in infinitely many different ways, the branch point is not. No matter how one decides to take the branch cut, the branch point will always be at the origin and it cannot be removed. Therefore, the massless mode cannot be a solution of Eq.  \eqref{eq:eomq}. Even if we try to go around this fact by shifting the pole at the origin by a small number $\varepsilon$
\begin{equation}
\oint_\mathcal{C}\mathrm{d}^4q\, e^{-iqx}(q-\varepsilon)^2\left[1-\frac{NG}{120\pi} q^2\log\left(-\frac{q^2}{\mu^2}\right)\right]\tilde h_{\mu\nu}(q) = 0,
\end{equation}
we would not be able to get consistent results. In fact, as a consequence of Sokhotski-Plemelj theorem, the limit when $\varepsilon\to 0$ does not commute with the integral sign, thus taking that limit in the end does not recover the theory in which $q_0=0$. This is just a manifestation of the well-known vDVZ discontinuity as $\varepsilon$ acts as a mass for the mode $q_0$.

\begin{figure}
\centering
\begin{subfigure}[b]{0.4\textwidth}
	\includegraphics[scale=0.5]{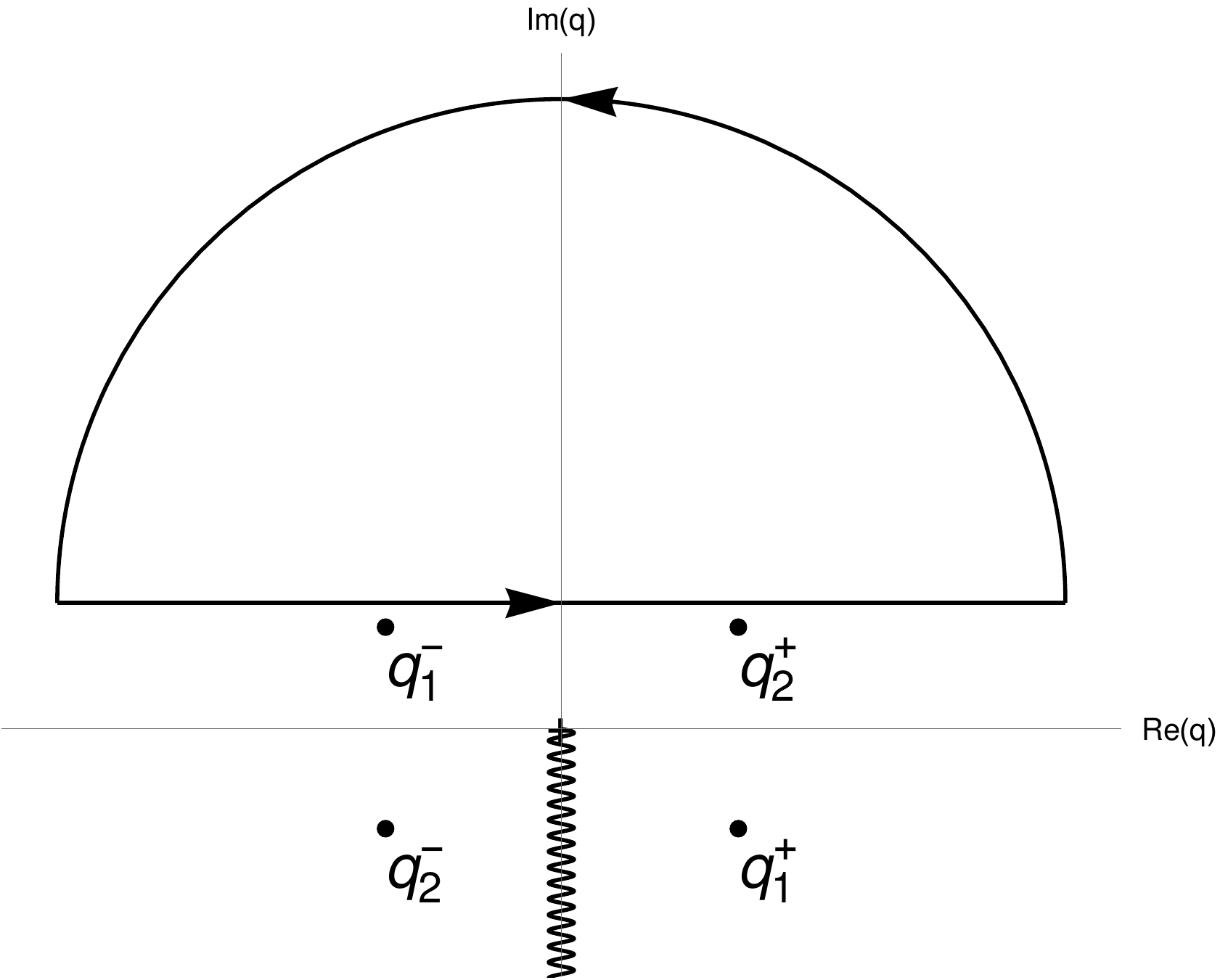}
	\caption{$t<0$}
	\label{fig:tneg}
\end{subfigure}
\quad\quad\quad
\begin{subfigure}[b]{0.5\textwidth}
	\includegraphics[scale=0.5]{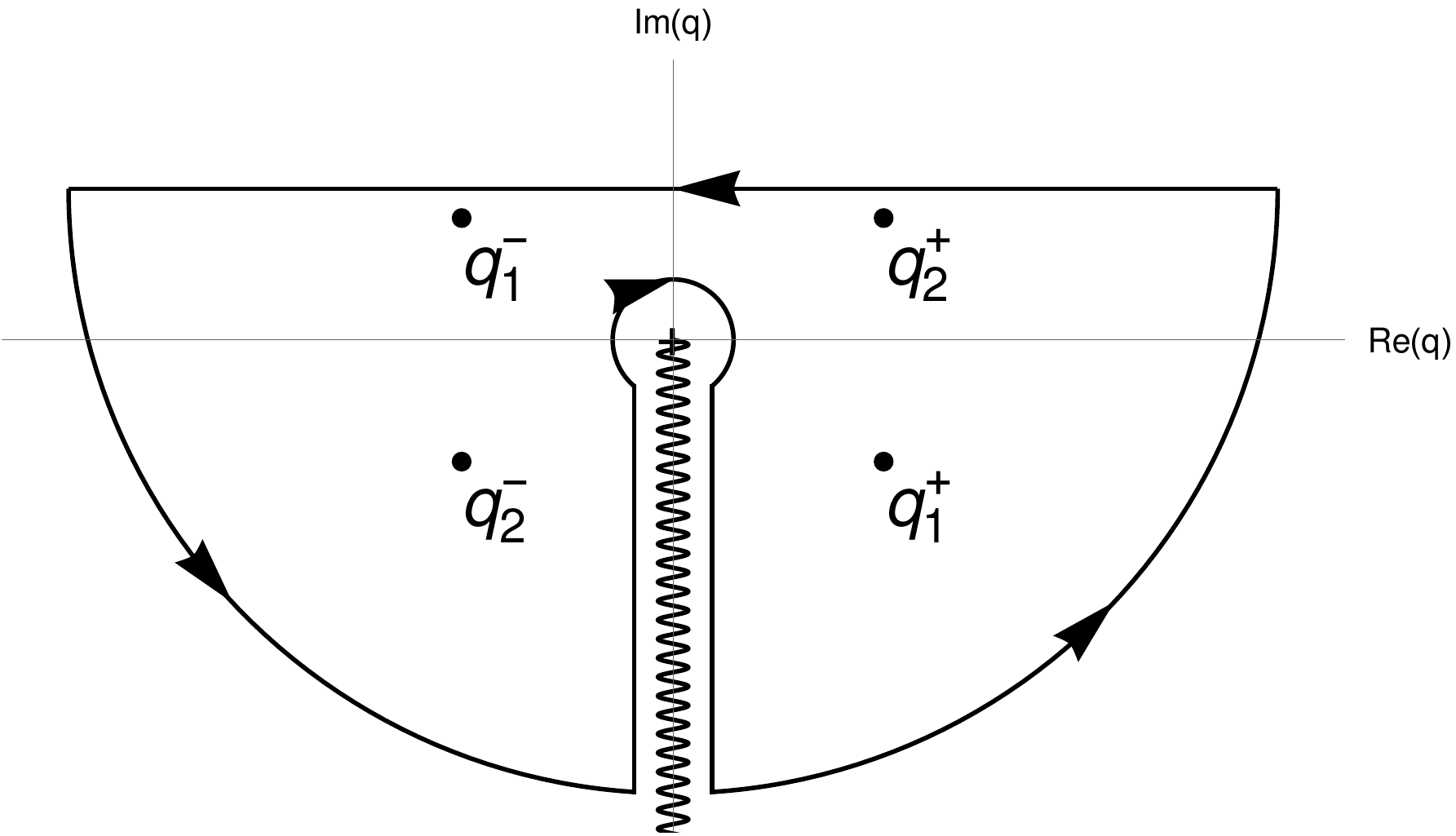}
	\caption{$t>0$}
	\label{fig:tpos}
\end{subfigure}
\caption{Contour $\mathcal{C}$ that ensures causality, enclosing no poles for negative times. The branch cut is represented by the wavy line.}
\label{fig:cont}
\end{figure}

Another thing to consider when choosing the contour $\mathcal{C}$ is causality. The quantum effective action \eqref{eq:nl} is obtained performing the standard in-out formalism. Thus its resulting propagator is the Feynmann one whose behaviour is acausal. This is not an issue when calculating scattering amplitudes, but should be taken care of when dealing with dynamical equations as in \eqref{eq:eom}. We shall enforce causality by properly choosing the contour $\mathcal C$ so that \eqref{eq:eom} satisfies retarded boundary conditions. This is equivalent to replacing the Feynmann propagator by the retarded one as performed in \cite{Barvinsky:1987uw,Vilkovisky:2007ny}. Figure \ref{fig:cont} shows the contour $\mathcal C$ that ensures a causal evolution. For $t<0$, the contour is chosen so that no poles are enclosed which yields a vanishing integral due to the Cauchy's integral theorem. For $t>0$, the contour encloses all poles while avoiding the branch cut.

Having chosen the contour of integration, we now need to find the $\tilde h_{\mu\nu}(q)$ that is able to cancel out all the zeros of $F(-q^2)$ that are enclosed by $\mathcal{C}$. But since $F(-q^2)$ is not a polynomial, the task of factorizing it becomes a bit more involved. Fortunately, according to Weierstrass factorization theorem\footnote{To avoid confusion, it is worth saying that this theorem only holds if $F(-q^2)$ is an entire function. We are actually using a generalization of this result to subsets of the complex plane where $F(-q^2)$ is holomorphic and which is attributed to Picard and Mittag-Leffler.}, $F(-q^2)$ can be decomposed as \cite{Ash}
\begin{equation}
F(-q^2) = q^2 e^{g(q^2)}\prod_{n=1}^{\infty}\left(1-\frac{q^2}{q_n^2}\right)\exp\left\{\left(\frac{q}{q_n}\right)^2+\frac{1}{2}\left(\frac{q}{q_n}\right)^4+\cdots+\frac{1}{\lambda_n}\left(\frac{q}{q_n}\right)^{2\lambda_n}\right\},
\label{eq:weier}
\end{equation}
where $g(q^2)$ is a holomorphic function, $\{\lambda_n\}$ is a sequence of integers and $q_n$ are the zeros of $F(-q^2)$. Therefore, the field solution must have the form
\begin{equation}
\tilde h_{\mu\nu}(q) = \sum_{n=1}^{2}\sum_{s=+,-} \frac{A^{n,s}_{\mu\nu}}{q-q_n^s},
\end{equation}
where $A^{n,s}_{\mu\nu}$ are constant tensors, to be able to cancel out the zeros of \eqref{eq:weier}. Transforming back to the position space gives
\begin{equation}
h_{\mu\nu}(x) = \sum_{n=1}^2\sum_{s=+,-} a^{n,s}_{\mu\nu} e^{-iq_n^sx},
\end{equation}
where $a^{n,s}_{\mu\nu}$ are polarization tensors. As it was previously observed in \cite{Calmet:2016sba}, the modes $q_3^-$ and $q_2^+$ have negative imaginary part that leads to a damping in these modes. The modes $q_2^-$ and $q_3^+$, on the other hand, have positive imaginary part, leading to an exponential growth. Thus, after a finite time the damped modes will die out and the behaviour of the fluctuation $h_{\mu\nu}$ will be dominated by the growing modes which will continue to grow indefinitely, eventually causing a gravitational collapse and destroying the whole structure of the spacetime. Note that the e-folding time is $\tau\sim \frac{\sqrt{N}}{M_p^2}$ and the time for which strong dynamics kicks in is $T\sim \tau\log(M_p^2\lambda^2)$, where $\lambda$ is the gravitational wavelength. Hence, even before the breakdown of the effective theory ($t<T$), the runaway modes are able to grow 80 orders of magnitude for a typical wavelength $\lambda\sim 10^3$km, thus one cannot simply assume that this problem is solved by strong dynamics above the Planck scale.

An obvious way to cure this issue is to choose the contour $\mathcal C$ without enclosing the poles $q_2^-$ and $q_3^+$. However, energy conservation requires that the other poles $q_2^+$ and $q_3^-$ are also removed. In fact, the energy lost by the damped modes is taken over in average by the runaway modes. In addition, CPT invariance requires that complex modes always arise in conjugate pairs. Therefore, the only way of consitently curing the instability of Minkowski spacetime is by removing the modes altogether via a suitable choice of the integration contour. Of course that this does not mean that the spectrum of the fundamental theory is empty as the particle spectrum should always be read off from the fundamental action and not from the effective one \cite{Belgacem:2017cqo}. The implication, however, is that there is no way to consistently perturb the theory without running into instabilities. The only leftover is the non-linear background equation of motion.

\section{Stability of de Sitter spacetime}
\label{sec:sitter}
We shall now turn our attention to the case where the background is curved. For an arbritrary curved background, this is highly non-trivial, but we can take advantage of the existence of symmetries in maximal-symmetric backgrounds in our favor. A good example of such background is de Sitter spacetime. It was shown that the de Sitter metric is also a solution of the background equation enhanced with quantum corrections \cite{Codello:2015pga}. Therefore, it makes sense to ask whether such background is stable under fluctuations of the metric in a quantum theory of gravity.

\begin{figure}
\centering
\includegraphics[scale=0.5]{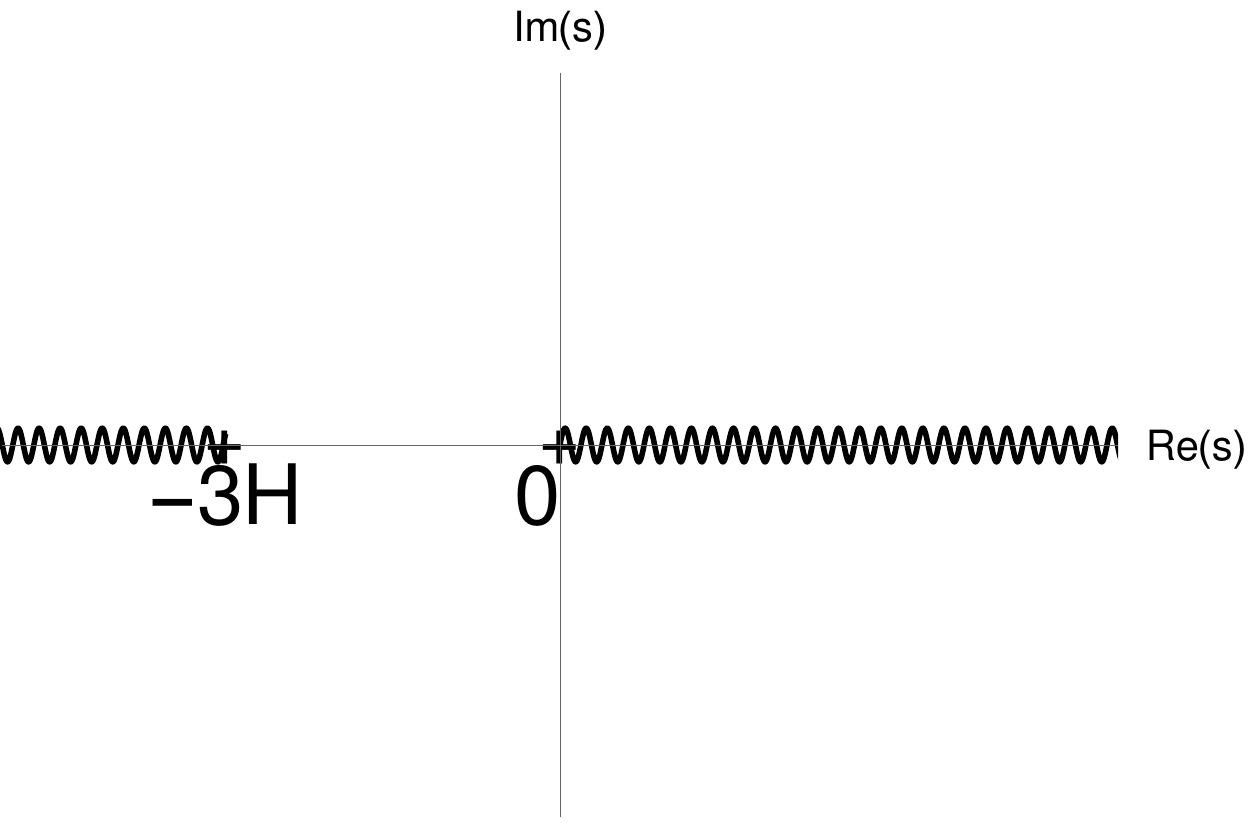}
\caption{Example of branch cuts starting at $s=-3H$ and at the origin $s=0$ and extending to the infinity.}
\label{fig:cuts}
\end{figure}

The de Sitter metric in the conformal coordinate reads
\begin{equation}
ds^2 = \frac{1}{H^2\tau^2}(d\tau^2 - dx_i^2),
\label{eq:confc}
\end{equation}
where $H$ is the Hubble constant, $\tau$ is the conformal time and $x_i$ denotes the three dimensional space. Let us assume that the fluctuations are homogeneous, thus $h_{\mu\nu}(x)=h_{\mu\nu}(\tau)$. The rescaled perturbation field $\tilde h_{\mu\nu}=H^2\tau^2h_{\mu\nu}$ in the transverse-traceless-synchronous gauge
\begin{equation}
\partial_i \tilde h_{ij} = 0,\qquad \tilde h_{0 \mu} = 0, \qquad\tilde h^i_{\ i} = 0,
\label{eq:tts}
\end{equation}
satisfies
\begin{align}
\Box_{\bar g} \tilde h_{ij} &= \frac{1}{\sqrt{-\bar g}}\partial_\mu\left(\sqrt{-\bar g}\bar g^{\mu\nu}\partial_\nu\tilde h_{ij}\right),\nonumber\\
&= (H\tau)^2\partial_\tau^2 \tilde h_{ij} - 2H^2\tau\partial_\tau \tilde h_{ij}.
\end{align}
Observe that the d'Alembert operator $\Box_{\bar g}$ is exactly the Laplace-Beltrami operator that acts on scalar fields even though it is being applied to a tensor field \cite{Grishchuk,Date:2015kma,Ford:1977dj,Rajaraman:2016nvv}. Although this is evident in the conformal patch \eqref{eq:confc} complemented with the gauge conditions \eqref{eq:tts}, it is useful to change the coordinates to an FLRW-like chart to make contact with cosmology:
\begin{equation}
ds^2 = dt^2 - e^{2Ht}dx_i^2.
\end{equation}
In these new coordinates, the d'Alembert operator reads
\begin{equation}
\Box_{\bar g} \gamma_{ij} = \partial_t^2 \gamma_{ij}+3H\partial_t \gamma_{ij},
\end{equation}
where $\gamma_{ij}(t) = \tilde h_{ij}(\tau(t))$. As the problem only involves time, it is convenient to use Laplace transform instead of Fourier's. Then, Eq. \eqref{eq:fourier} turns into
\begin{equation}
\gamma_{ij}(x) = \oint_\mathcal{C}\mathrm{d}s\, e^{st}\gamma_{ij}(s)
\end{equation}
and Eq. \eqref{eq:eom} now becomes
\begin{equation}
\oint_\mathcal{C}\mathrm{d}s\, e^{st}(s^2+3Hs)\left[1+\frac{NG}{120\pi}(s^2+3Hs)\log\left(\frac{s^2+3Hs}{\mu^2}\right)\right]\gamma_{ij}(s) = 0.
\end{equation}
This time, two branch points show up coinciding with the zeros $s_0=0$ and $s_1=-3H$. The corresponding branch cuts start at $s_0$ and $s_1$ and goes up to infinity (Figure \ref{fig:cuts}), preventing us from including the massless mode $s_0$ and the Hubble friction $s_1$ in the theory. Therefore, $\gamma_{ij}$ must only develop poles at
\begin{align}
s_2^\pm &= -\frac{3H}{2}\left(1\pm\sqrt{1-\frac{4q_2^2}{9H^2}}\right),\\
s_3^\pm &= -\frac{3H}{2}\left(1\pm\sqrt{1-\frac{4(q_2^2)^*}{9H^2}}\right),
\end{align}
where $q_2$ is the pole \eqref{eq:q2} of Minkowski spacetime. Damped (growing) modes have negative (positive) real part, thus for the renormalization scale \eqref{eq:scale} $s^-_2$ and $s^-_3$ are damped while $s^+_2$ and $s^+_3$ are exponentially growing, showing that de Sitter spacetime is also unstable in quantum gravity.

As before, we could choose a contour to eliminate the runaway modes, but because of energy conservation and CPT invariance, we must also eliminate the damped ones. However, in a de Sitter background there is a more attractive alternative solution where the contour encloses all the poles but the ones over branch points. In the limit of large Hubble constant $|q_2|\sim \frac{M_p}{\sqrt N}\ll H$, both $s^\pm_2$ and $s^\pm_3$ approximately collapses into the branch points $s_0$ and $s_1$, solving the instability issue. Note, however, that this is an approximation as $|q_2|/H$ is not identically zero. Therefore, we are still able to choose a contour that encloses $s^\pm_2$ and $s^\pm_3$ without crossing the branch cuts. This way, we end up with a richer and more interesting theory than the one with no modes. It is important to stress that the instability is automatically solved in the limit of a large number of fields $N$. Coincidentally, this limit is also required for unitarity \cite{Aydemir:2012nz}.

From a cosmology viewpoint, de Sitter is just an idealization of the more realistic situation where the Hubble constant is time-dependent as described by a general FLRW model. Thus, if the Hubble constant gets large for large times, its dynamics could trigger the existence of the growing modes during a finite time $t_0$ in a controlled manner, without distabilizing the spacetime and leading to important implications for astrophysics. Mathematically, in order to prevent the runaway modes from growing indefinitely, the Hubble constant evolution must satisfy
\begin{equation}
H(t)\gg |q_2|,\quad t>t_0,
\end{equation}
for some reference time $t_0$. This way, when the growing modes start to get too large, the Hubble constant will also become large, turning the growing modes $s^+_{2,3}$ into damped modes and the damped modes $s^-_{2,3}$ into constants. In this situation, the field solution is
\begin{align}
\gamma_{ij}(t) &= \sum_{n=1}^2\sum_{k=+,-} a^{n,k}_{ij} e^{s_n^kt}\\
&\approx \left(a^{1,+}_{ij}+a^{2,+}_{ij}\right)e^{-3Ht} + a^{1,-}_{ij}+a^{2,-}_{ij},\quad\quad\quad\quad (H\gg |q_2|),
\label{eq:fsol}
\end{align}
which is constant for large times. Note that Eq. \eqref{eq:fsol} does not violate energy conservation as the decaying exponential exists only due to the Hubble friction.

\section{Conclusions}
\label{sec:conc}
We showed how quantum gravitational effects lead to instabilities in Minkowski and de Sitter backgrounds. Even before reaching the scale where the effective theory breaks down, the exponentially growing modes are able to increase by 80 orders of magnitude, which is more than enough to cause the whole universe to collapse gravitationally. This situation is obviously unphysical as we would have observed gravitational waves with huge amplitudes by now (most likely long before LIGO). In the Minkowski background, we solved this problem by selecting a contour that does not enclose the runaway modes. For physical reasons, however, we were forced to remove the modes altogether. In the de Sitter spacetime, a more interesting alternative was given for when the Hubble constant satisfies $H\gg\tfrac{M_p}{\sqrt N}$. We argued that an FLRW universe, being a more realistic version of de Sitter, could provide an interesting solution to the instability problem by triggering the existence of the runaway modes in a controlled way, keeping them from growing indefinitely.

\noindent{\it Acknowledgments:}
 I. K. is supported by the National Council for Scientific and Technological Development  -- CNPq (Brazil) under Grant No. 155342/2018-5. RdR~is grateful to FAPESP (Grant No.  2017/18897-8) and to  CNPq (Grant No. 303293/2015-2), for partial financial support.


\bigskip{}

\end{document}